# Queue Model of Leaf Degree Keeping Process in Gnutella Network

Chunxi Li and Changjia Chen

School of Electrical and Information Engineering
Beijing Jiaotong University
Beijing 100044, P. R. China
86-10-51684759

cxl@telecom.njtu.edu.cn,   changjiachen@sina.com

*Abstract*—Leaf degree keeping process of Gnutella is discussed in this paper. Queue system based on rules of Gnutella protocol are introduced to modeling this process. The leaf degree distributions resulted from the queue system and from our real measurement are compared. The well match of those distributions reveal that the leaf degree distribution in Gnutella network should not be power law or power law like as reported before. It is more likely a distribution driven by certain queue process specified by the protocol.

## I. INTRODUCTION

As an important p2p application, Gnutella has drawn the attention of many researchers. Up to our knowledge there have been two important measurements [1,2] about this network and both of them reported a power law or power law like distribution in peer degree distributions. Our recent work [3] based on a continue watching suggests to break the peer degree into ultra degree (*Udeg*) and leaf degree (*Ldeg*) that indicate how many ultra peers and leaf peers are currently connected respectively by a given peer. Phase space representation is introduced to study the peer distributions as leaf degree distribution and ultra degree distribution separately and jointly. Totally different distributions on Ldeg and Udeg are reported based on our measurement. A simple M/M/m/m loss queue is introduced to model this degree keeping process. In this paper we will refine our previous work in modeling the degree keeping process. We will show that, characters in our reported distribution are rooted in the protocol features of the network. A queue system that mimics the main feature of the protocol will generate very similar degree distribution as we observed in our measurement.

Our result in this paper is useful in interpreting the measurement results, in theoretical study the Gnutella network, and in simulations to generate the more realistic degree traces.

## II. PROTOCOL FEATURES IN LEAF CONNECTION OF ULTRAPEER

In this paper, only LimeWire peers are considered since that kind of peers dominet the network and the LimeWire implementation are best documented publicly among other software versions. It is reported [4] that

"Overall, the ratio between LimeWire and BearShare has been fairly stable, with LimeWire making up 75–85% of ultrapeers, BearShare making up 10–20%, and other brands making up 3–7%."

It is well known that Gnutella is a two-tier network with ultra peers as the core of the network. In fact only the degrees of ultra peers are interested since a leaf peer only connects to very few (2-3) ultra peers. In this paper we will concentrated on the modeling of the leaf degree keeping process of LimeWire ultra peers. In the following paper we will study the ultra degree keeping process of LimeWire ultra peers.

There are several features listed in documents of LimeWire about the rules that how an ultra peer connects to the leaf peers:

"As a LimeWire ultrapeer, we have 32 slots for ultrapeers and 30 slots for leaves. If a leaf connects, and we have an open slot for it among our 30, we accept it." [5], "LimeWire ultrapeers will allow ANY leaf to connect, so long as there are at least 15 slots open. Beyond that number, LimeWire will only allow 'good' leaves. To see what constitutes a good leave, view HandshakeResponse.isGoodLeaf(). To ensure that the network does not remain too LimeWire-centric, it reserves 3 slots for non-LimeWire leaves."[6]. "GOOD_LEAF = GOOD_ULTRAPEER && (IS_LIMEWIRE || NO_REQUERYING)",[7] "good ultrapeers or just those who support UDP pinging" [8]

In summary there are three key parameters in guiding the leaf connection of an ultra peer: maximum connections $C_m$=30, Good leaf threshold $C_g$=15 and Non LimeWire reserve $C_n$=3. An ultra peer will never admit more than $C_m$ leaf connections. If the admitted leafs exceed $C_g$ only good leaf will be admitted. There are 3 slots reserved for non-LimeWire leafs.

## III. THE LIFE OF ULTRA PEER

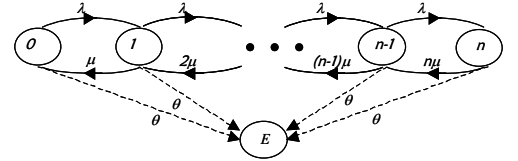

Fig 1. M/M/m/m queue system

In [3] we model the leaf degree keeping process as an M/M/m/m loss queue system in figure 1 (without the dish line). Where state *k* indicates the ultra peer has admitted k leaf connections currently. In other word the ultra peer has Ldeg of *k*. Where $\lambda$ is the arriving rate of leaf connection and $\mu$ is the dropping rate of single leaf connection. The state transition matrix $A$ for this system is

$$A_{\lambda,\mu} = \begin{pmatrix} -\lambda & \mu & 0 & 0 & 0 \\ \lambda & -\lambda-\mu & 2\mu & \vdots & \vdots \\ 0 & \lambda & -\lambda-2\mu & \cdots & 0 \\ \vdots & \vdots & \cdots & \cdots & n\mu \\ 0 & 0 & & \lambda & -\lambda-n\mu \end{pmatrix}$$

The equilibrium distribution $q$ satisfying $A_{\lambda,\mu}q$=0 and can be solved as $q_i=\alpha(\lambda/\mu)^i/i!$. The best fitting equilibrium distribution $q_{IL}$ (the line with blue *) and the real measured distribution $q_{RM}$ (the line with red square) are shown in figure 2. In this paper, the best fitting is always in the $L_1$ distance and the $L_1$ distance for two vectors $p$, $q$ is defined as $d(p,q)=\Sigma_i|p_i-q_i|$. In the best fitting we have $\lambda/\mu$=48.7979 with a fitting error $\Delta=d(q_{IL}, q_{RM})$= 0.2456 . If $\mu=0.21$(in the time scale of 0.5 hour as in our measurement), then $\lambda=10.2476$. Look at the log plot (the right subplot of figure 2), one can find the fitting error is mainly in the region of small degrees (0-20). Based on our measurement data [3], the average life of leaf connection is about 2.4 hours and the average life of ultra peer is about 10 hours. We believe that a finite lived M/M/m/m queue system as shown in figure 1 include the dashed lines can fit the degree keeping process more accurately. In this system, $\theta$ is the death rate of an ultra peer. The average life of 10 hours of an ultra peer corresponds to $\theta$ =0.052 in a time scale of half an hour. Let $e_0=(1, 0, …, 0)^T$ be the vector with 1 at its first element and zeros for all other elements. The equilibrium distribution $q_{FL}$ for the finite life system satisfying $q_{FL}=\alpha(A_{\lambda,\mu}-\theta I)^{-1}e_0$. This can be proved straightforward. Let $q(t)=(q_i(t), i=0,…,n)$, where $q_i(t)$ is the probability of state $i$ at time $t$ with initial state distribution $e_0$. Based on the diagram in fig 1, we have the differential equations $q'(t)=(A_{\lambda,\mu}-\theta I)q(t)$. This equation can be easily solved as $q(t)=\exp\{(A_{\lambda,\mu}-\theta I)t\}e_0$. The equilibrium distribution is defined as normalized $q_{FL}=\int_0^\infty q(t)dt$ . Hence we have $q_{FL}=\alpha(A_{\lambda,\mu}-\theta I)^{-1}e_0$.

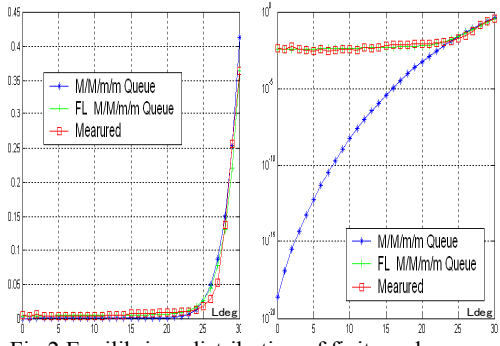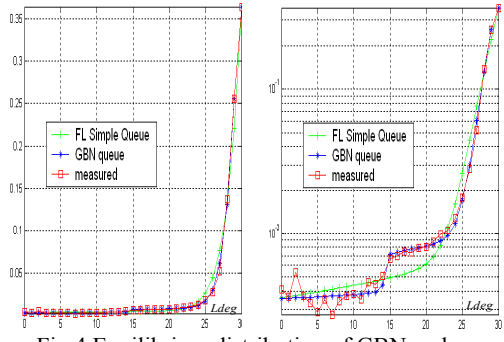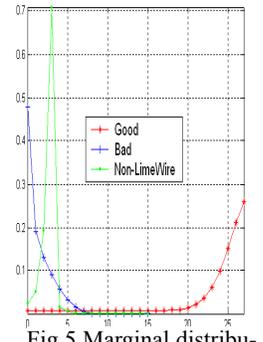

Fig 2 Equilibrium distribution of finite and infinite life queue vs. measured distribution

Fig 4 Equilibrium distribution of GBN and Simple queue vs. measured distribution

Fig 5 Marginal distributions of GBN queue

The best fitting $q_{FL}$ (the line with green +) is also shown in figure 2. It has a fitting parameters $\lambda$=10.5179, $\mu$= 0.2116, $\theta$ =0.0374 and a fitting error $\Delta=d(q_{FL}, q_{RM})$=0.1185. The fitting parameters are close to the real measured values and the fitting error is significantly smaller than that of infinite life system, especially at small degree region. Based on above discussions, we conclude that the finite life of ultra peer must be taken in consideration in modeling the degree keeping process of Gnutella. In the following of this paper, only finite life systems are being discussed.

### IV. THE QUEUE SYSTEM WITH FULL FEATURES OF THE PROTOCOL

The full protocol features we considered here are: (1) Finite ultra peer life; (2) Maximum leaf connections $C_m$; (3) Good leaf threshold $C_g$; (4) Non-LimeWire reserve $C_n$. We will name the corresponding queue system as GBN queue system. A GBN queue system has a state $\{(k_g, k_b, k_n): k_g+k_b+k_n \leq C_m$ and $k_g+k_b \leq C_m-C_n=C_{gb}$ and $k_b, k_n \leq C_g$ and $k_g \leq C_{gb}\}$ where $k_g$, $k_b$ and $k_n$ stand for the number of good leaf LimeWire connections, bad leaf LimeWire connections and non-LimeWire connections respectively a peer connected currently. In a GBN queue we will use $\lambda_g$, $\lambda_b$, and $\lambda_n$ ($\mu_g$, $\mu_b$, and $\mu_n$) to represent arrive (departure) rates of good, bad and non-LimeWire connections respectively. These connection rules can be translating into a set of state transition rules as depicted in figure 3. In figure 3, $\delta_{\pi(x)}$ is the indicating function of statement $\pi(x)$. The value of $\delta_{\pi(x)}$ is 1 if $\pi(x)$ is true and 0 otherwise. There a three classes of rules in figure3: connection arriving, connection departure and life termination. The life termination rate is $\theta$ as we have discussed in previous section. The connection departure rules are very much like that of M/M/m/m queue we have discussed before. The constraints here are simply exclusion of boundary states. For example there is no connection drop at state 0 and the up boundary state can be reached only by admitting of new connection. The protocol features are mainly reflected in the connection arriving rules. For a given state ($k_g$, $k_b$, $k_n$), a good leaf LimeWire connection can be admitted iff there is not "*too LimeWire-centric*" ($k_g+k_b<C_m-C_n=C_{gb}$) and there is room for new connection ($k_g+k_b+k_n<C_m$); A bad leaf LimeWire connection can be admitted iff the good leaf threshold has not be exceeded ($k_g+k_b+k_n<C_g$); A non-LimeWire connection can be admitted iff the Non-LimeWire reserve has not be exceeded ($k_n<C_n$) or the Non-LimeWire reserve is exceeded but the good leaf threshold has not be exceeded ($k_n \geq C_n$ and $k_g+k_b+k_n<C_g$). A set of differential equations following these rules can be written as:

$$p'_{k_g,k_b,k_n} = -(\lambda_g \delta_{(k_g+k_b<C_{gb} \wedge k_g+k_b+k_n<C_m)} + \lambda_b \delta_{k_g+k_b+k_n<C_g}$$
$$+ \lambda_n \delta_{\{(k_n<C_n) \vee (k_n \geq C_n \wedge k_g+k_b+k_n<C_g)\}} + k_g \mu_g \delta_{k_g>0} + k_b \mu_b \delta_{k_b>0} + k_n \mu_n \delta_{k_n>0} +$$
$$+ \theta)p_{k_g,k_b,k_n} + \lambda_g \delta_{k_g>0} p_{k_g-1,k_b,k_n}(t) + \lambda_b \delta_{(k_b>0) \wedge (k_g+k_b+k_n \leq C_g)} p_{k_g,k_b-1,k_n} +$$
$$+ \lambda_n \delta_{\{(0<k_n \leq C_n) \vee (C_g>k_n>C_n \wedge k_g+k_b+k_n \leq C_g)\}} p_{k_g,k_b,k_n-1} +$$
$$(k_g+1)\mu_g \delta_{(k_g+k_b<C_g \wedge k_g+k_b+k_n<C_m)} p_{k_g+1,k_b,k_n} +$$
$$+ (k_b+1)\mu_b \delta_{(k_b<C_g \wedge k_g+k_b<C_{gb} \wedge k_g+k_b+k_n<C_m)} p_{k_g,k_b+1,k_n}$$
$$+ (k_n+1)\mu_n \delta_{(k_n<C_g \wedge k_g+k_b+k_n<C_m)} p_{k_g,k_b,k_n+1}$$

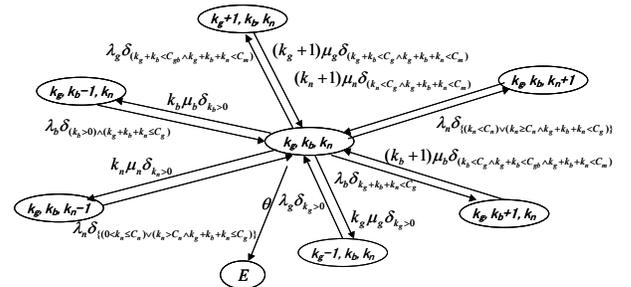

Fig 3. State Transition Rules of GBN Queue

Above equation can also be written into a matrix form as $q'(t)=(A_{\lambda,\mu}-\theta I)q(t)$ as we did before, and has a equilibrium distribution $q_{GBN}=\alpha(A_{\lambda,\mu}-\theta I)^{-1}e_0$. A MatLab program is written to find the best fitting equilibrium distribution $q_{GBN}$ to the real measured distribution $q_{RM}$. The resulted $q_{GBN}$ is drawn in figure 4 along with real measured distribution $q_{RM}$ and the $q_{FL}$ the best fitting of finite life M/M/m/m queue in figure 2. The $q_{GBN}$ fits much better than $q_{FL}$. The best fitting is reached at $\lambda_g$=11.0926, $\lambda_b$=5.6722, $\lambda_n$=3.5248, $\mu_g$=0.1824, $\mu_b$=0.1828, $\mu_n$=0.2980, and $\theta$=0.0714 with a fitting error $\Delta=d(q_{GBN}, q_{RM})$= 0.0306. Roughly we can use $\lambda_n/(\lambda_g+\lambda_b)$= 0.2103 to estimate the ration of non-LimeWire and LimeWire peers. It is similar to the reported ration in [4]. Up to our knowledge no one has ever measured the good leaf and bad leaf massively. Based on our fitting result, the life of good and bad leaf connections is similar and is about 1/3 of the life of ultra peers. But the life of non-LimeWire leaf connection is about 1/4 of the life of ultra peers. The marginal distributions for good, bad LimeWire and non-LimeWire connections are depicted in figure 5. The means of good, bad LimeWire and non-LimeWire connections are 23.03, 1.26 and 2.66 respectively. We are trying to verifying above estimations through the measurement of real network.

## V. SIMPLIFIED APPROXIMATE MODEL

There are 4000 states in the GBN queue model. The calculations are tremendous in solve the best fitting problem of GBN queue, since a 4000x4000 matrix must be inversed at each iteration steps. For decreasing the number of states, we can think there are only two classes of connections: LimeWire and non-LimeWire. Instead of using 3 classes of connections we can adjust arriving and departure rate according to the good leaf threshold. There is an arriving and departure rate

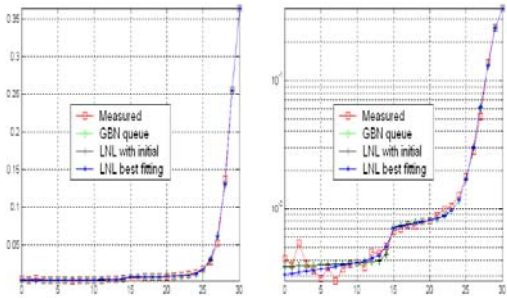

Fig 6 Equilibrium distribution of LNL vs. GBN and measured distribution

under the good leaf threshold and another arriving and departure rate above the good leaf threshold. We call this model as LNL model. A LNL queue system has a state $\{(k_l, k_n): k_l+k_n\leq C_m$ and $k_l \leq C_m-C_n=C_{gb}$ and $k_n\leq C_g\}$ where $k_l$ and $k_n$ stand for the number of LimeWire connections and non-LimeWire connections respectively. We will use $\lambda_a$, $\lambda_b$, and $\lambda_n$ ($\mu_a$, $\mu_b$, and $\mu_n$) to represent arrive (departure) rates of above good leaf threshold, below good leaf threshold and non-LimeWire connection respectively. For a given state $(k_l, k_n)$, $(\lambda_a, \mu_a)$ is applied to state $k_l$ if it is above good leaf threshold $(k_l+k_n\geq C_g)$. Otherwise $(\lambda_b, \mu_b)$ is applied. A set of differential equations following these rules can be written as:

$p'_{k_l,k_n} = -(\lambda_a\delta_{(k_l<C_{gb}\wedge k_l+k_n<C_m\wedge k_l+k_n\geq C_g)} + \lambda_b\delta_{k_l+k_n<C_g} +$
$+ \lambda_n\delta_{\{(k_n<C_n)\vee(C_g>k_n\geq C_n\wedge k_l+k_n<C_g)\}} + k_l\mu_a\delta_{k_l+k_n\geq C_g} + k_l\mu_b\delta_{(k_l>0\wedge k_l+k_n<C_g)} +$
$+ k_n\mu_n\delta_{k_n>0} + \theta)p_{k_l,k_n} + \lambda_a\delta_{(k_l>0\wedge k_l+k_n>C_g)}p_{k_l-1,k_n}(t) +$
$\lambda_b\delta_{(k_b>0\wedge k_l+k_n\leq C_g)}p_{k_l-1,k_n} + \lambda_n\delta_{\{(0<k_n\leq C_n)\vee(k_n>C_n\wedge k_l+k_n\leq C_g)\}}p_{k_l,k_n-1} +$
$+ (k_l+1)\mu_a\delta_{(k_l<C_{gb}\wedge k_l+k_n<C_m\wedge k_l+k_n\geq C_g)}p_{k_g+1,k_n}$
$+ (k_l+1)\mu_b\delta_{(k_l+k_n<C_g)}p_{k_l+1,k_n} + (k_n+1)\mu_n\delta_{(k_n<C_g\wedge k_l+k_n<C_m)}p_{k_l,k_n+1}$

Only the inverse of a 370x370 matrix is involved in fitting LNL model much simpler than that of GBN. Let ($\lambda'_g$, $\lambda'_b$, $\lambda'_n$, $\mu'_g$, $\mu'_b$, $\mu'_n$, $\theta'$) be the fitting parameter of GBN, then we chose a initial value for LNL as ($\lambda_a=\lambda'_g$, $\lambda_b=\lambda'_g+\lambda'_b$, $\lambda_n=\lambda'_n$, $\mu_a=\mu'_g$, $\mu_b=\mu'_b$, $\mu_n=\mu'_n$, $\theta=\theta'$ ). The best fitting equilibrium distribution $q_{LNL}$ (blue star line) and the distribution directly calculated from the initial value ( black + line) are drawn in figure 6 along with real measured distribution $q_{RM}$ and the $q_{GBN}$. The best fitting is reached at $\lambda_a$=11.1363, $\lambda_b$=19.7656, $\lambda_n$=3.5906, $\mu_a$=0.1849, $\mu_b$=0.5330, $\mu_n$=0.2984, and $\theta$=0.0707 with a fitting error $\Delta=d(q_{LNL}, q_{RM})$=0.0301. The error of the LNL distribution directly calculated from the initial value is $\Delta=d(q_{LNL}, q_{RM})$= 0.03067.

## VI. CONCLUSION

Only leaf degree distribution is discussed in this paper. It seems that the ultra degree distribution cannot be modeled simply based on the interaction between ultra peers. Leaf degree keeping process must be considered in modeling ultra degree keeping process. We are working on this problem now.

Through modeling of the leaf degree keeping process as a queue system, we show that our measured leaf degree distribution can be fitted very well. Since the queue modeling is an abstracted representation of the real protocol rules, we wish to persuade that it is more reasonable to believe the degree distribution of Gnutella has a shape like our measured result rather than to think it is as a power law or power law like.

REFERENCES

[1] Stefan Saroiu, P. Krishna Gummadi, and Steven D. Gribble, "Measuring and Analyzing the Characteristics of Napster and Gnutella Hosts," *Multimedia Systems Journal*, vol. 8, no. 5, Nov. 2002.
[2] Daniel Stutzbach, Reza Rejaie, and Subhabrata Sen. "*Characterizing Unstructured Overlay Topologies in Modern P2P File-Sharing Systems*", Internet Measurement Conference, October, 2005.
[3] Chunxi Li and Changjia Chen, "Gnutella: Topology Dynamics On Phase Space", submitted to IEEE/ACM Transactions on Networking, Spt., 2006.
[4] Amir H. Rasti, Daniel Stutzbach, Reza Rejaie, "On the Long-term Evolution of the Two-tier Gnutella Overlay," http://www.cs.uoregon.edu/~reza/PUB/gi-2006-long-term.pdf
[5] From http://www.limewire.org/wiki/index.php?title=Ultrapeers
[6] From http://www.limewire.org/nightly/javadocs/com/limegroup/gnutella/ConnectionManager.html
[7] From http://www.limewire.com/english/content/download.shtml
[8] From http://www.limewire.org/nightly/clover/results/com/limegroup/gnutella/messages/vendor/UDPCrawlerPong.html